# XRISM Quick Reference

xrism.isas.jaxa.jp

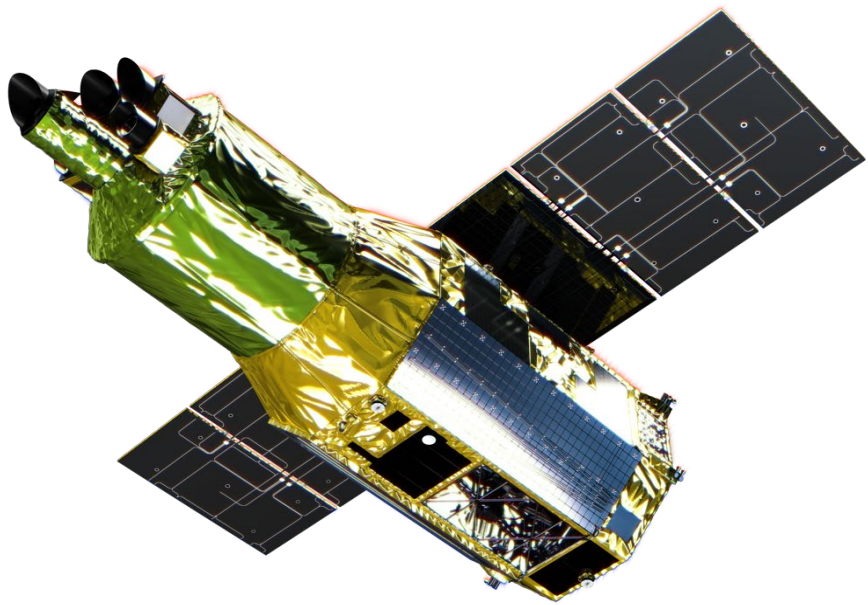

XRISM SOC-PVO

# Releases

- 2202-02-22 Version 2.1 (N. Ota, T. Mizuno, H. Uchiyama, the XRISM SOC-PVO team)
- 2022-02-04 Version 2 (N. Ota, T. Mizuno, H. Uchiyama, the XRISM SOC-PVO team)
- 2020-10-06 Version 1 (N. Ota, T. Mizuno, H. Uchiyama, the XRISM SOC-PVO team)

# Table of Contents

# I. Overview
# I-1. XRISM Mission

## Overview

The X-ray Imaging and Spectroscopy Mission (XRISM) is an X-ray observatory, which is the 7th in the series of the X-ray observatories from Japan. The mission of XRISM is to recover and resume the study of the prime objective of ASTRO-H/Hitomi "to solve outstanding astrophysical questions with high-resolution X-ray spectroscopy". It is currently planned to be launched in FY2022 with an HII-A rocket from the Tanegashima Space Center, Kagoshima, Japan.

| Launch site | Tanegashima Space Center |
|---|---|
| Launch vehicle | JAXA HII-A rocket |
| Orbit altitude | 550±50 km |
| Orbit type | Approximately circular orbit |
| Orbit inclination | 31 degree |
| Dimension | 7.9 m x 9.2 m x 3.1 m |
| Mass | 2.3 metric ton |
| Mission life | 3 years + cryogen free operation |

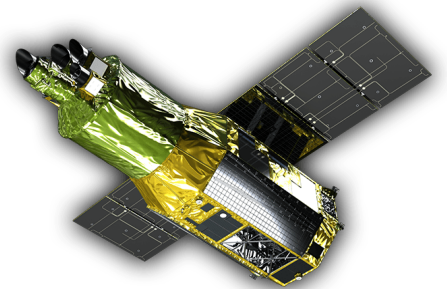

**Fig. 1** *Artist's drawing of the XRISM satellite.*

## Scientific objectives

- Revealing the structure formation of the Universe and evolution of clusters of galaxies
- Understanding the circulation history of baryonic matters in the Universe
- Investigating the transport and circulation of energy in the Universe
- Realizing the new science with high-resolution X-ray spectroscopy

## Key technologies

- An X-ray micro-calorimeter detector which enables high resolution (≤ 7 eV) spectroscopic observations between 0.3 and 12 keV
- An X-ray imager detector which enables a wide-field imaging spectroscopy between 0.4 and 12 keV

# I. Overview
# I-2. Spacecraft & Components

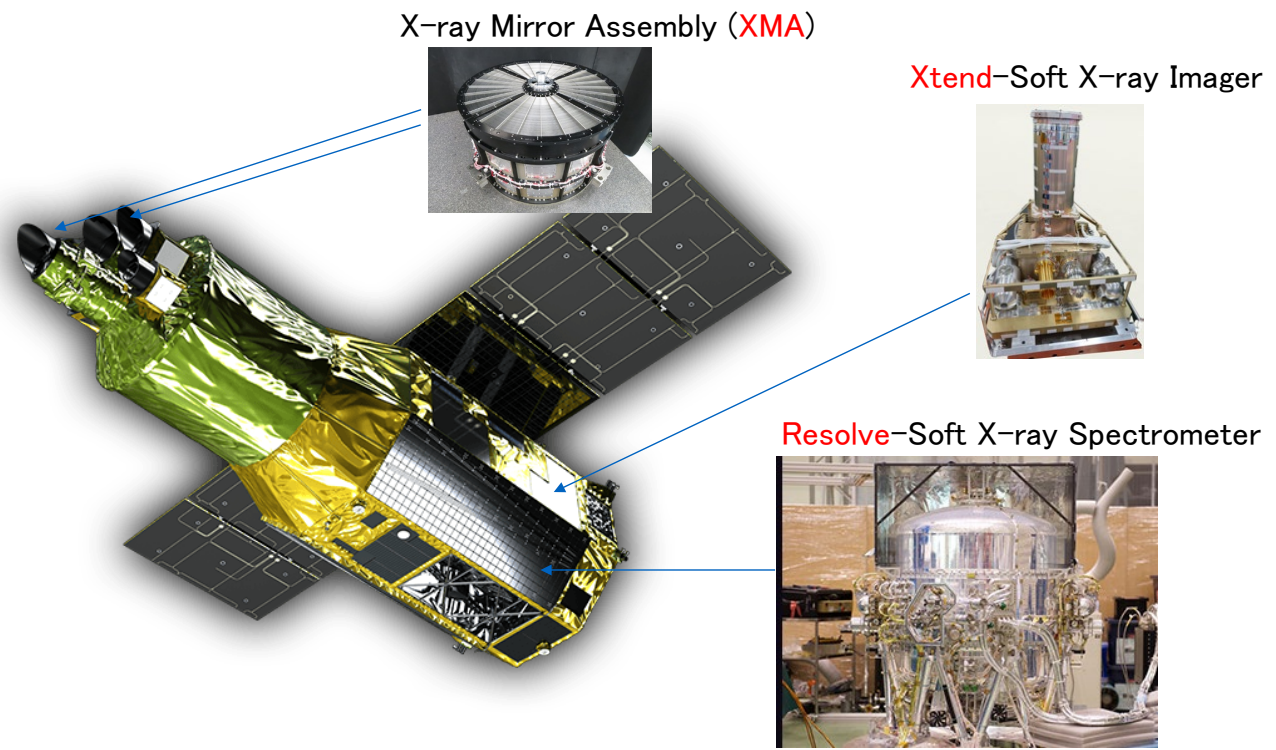

*Fig. 2* Schematic view of the XRISM satellite. There are two focusing X-ray telescopes mounted at a fixed optical bench. They will focus X-rays onto focal plane detectors mounted on the base plate of the spacecraft.

# I. Overview
# I-3. Parameters

*Table 1* *Key parameters and performance requirement of the XRISM observatory (Tashiro et al. 2018)*

| Parameters | Resolve | Xtend |
|---|---|---|
| X-ray mirrors | Conically approximated Wolter I optics (203 nested) | |
| Focal length | 5.6 m | |
| Angular resolution | ≤ 1.7 arcmin (HPD[*1]) | |
| Detector technology | X-ray micro-calorimeter | X-ray CCD |
| Effective area | ≥210 cm$^2$ @ 6keV, ≥160 cm$^2$ @ 1keV | ≥300 cm$^2$ @ 6 keV |
| Field of View | ≥ 2.9 x 2.9 arcmin$^2$ | ≥ 30 x 30 arcmin$^2$ |
| Energy range | 0.3 – 12 keV | 0.4 – 12 keV |
| Absolute energy scale | ≤ 2 eV | – |
| Energy resolution | ≤ 7 eV FWHM @ 6keV | ≤ 250 eV @ 6keV (EOL) |
| Non X-ray background | ≤ 2 x 10$^{-3}$ c/s/keV/array | ≤ 1 x 10$^{-6}$ c/s/keV/arcmin$^2$ (in 5–10 keV) |
| Time tagging accuracy | ≤ 1 ms | – |

*1 Half Power Diameter

## II. Instruments

# II-1. X-ray Mirror Assembly

### Basics

XRISM has two identical X-ray Mirror Assembly (XMA). One is for Resolve and the other for Xtend. XMA is similar in concept to the X-ray telescope (XRT) of Suzaku (**Table 2**). It is composed of 203 thin reflector shells tightly nested confocally and coaxially (**Fig. 3**). A conical approximation of Wolter-I optic is used.

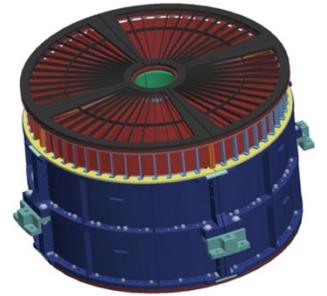

*Fig. 3* Schematic view of XMA. The blue and red parts are the mirror housing and Pre-Collimator, respectively.

*Table 2 Comparison of XRISM/XMA and Suzaku/XRT*

|       | No. of tele scope | Focal length (m) | Diameter (cm) | No. of nested shells | HPD (arcmin) | $A_{eff}^{*1}$ (cm$^2$) |
|-------|---|------|----|-----|-----|-----------|
| XMA   | 2 | 5.6  | 45 | 203 | 1.3 | 560 / 425 |
| XRT-I | 4 | 4.75 | 40 | 175 | 2.0 | 470 / 320 |

$^{*1}$ $A_{eff}$ is the on-axis effective area (mirror only) at 1 & 6 keV. Source extraction radius is infinite, or 100% photons are encircled. The values correspond to one unit.

### Technology

Three different thicknesses of Al substrates (152, 229, and 305 μm) are used, in which the outer shells use the thicker substrates. This is intended for achieving a large collecting area with a better imaging quality than Suzaku/XRT. The considerably better HPD than XRT (**Table 2**) is realized by fixing the reflectors to support bars with adhesive. Stray light or contaminating X-rays from outside of the field of view, is reduced by a stray-light baffle, called Pre-Collimator, which consists of coaxially-nested cylindrical aluminum blades placed above each reflector. A thermal shield is attached in front of the Pre-Collimator to stabilize the thermal environment of the XMA. The shield is made of Al-coated Polyimide film to ensure a large effective area in the soft energy band.

### Performance

The XMA's effective area at on-axis and off-axis angles and the encircled effective area are shown in **Fig. 4** and **Fig. 5**, respectively.

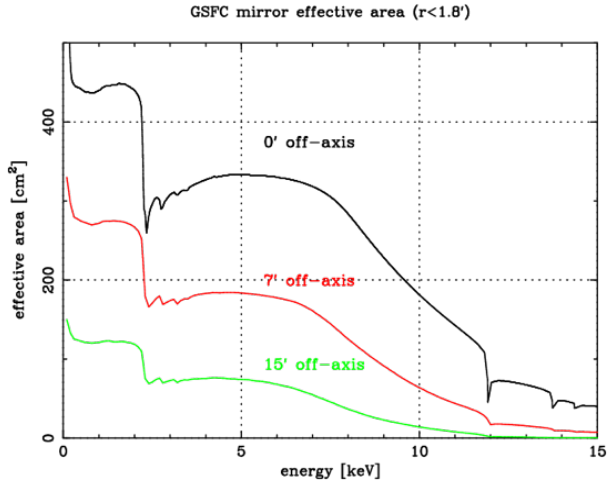

*Fig. 4* Effective area (mirror only) as a function of energy at on-axis and some off-axis angles. The source extraction radius is 1.8 arcmin (78% encircled energy fraction). Note that this is a mirror-only effective area. **Fig. 14** shows the actual effective areas combined with detectors.

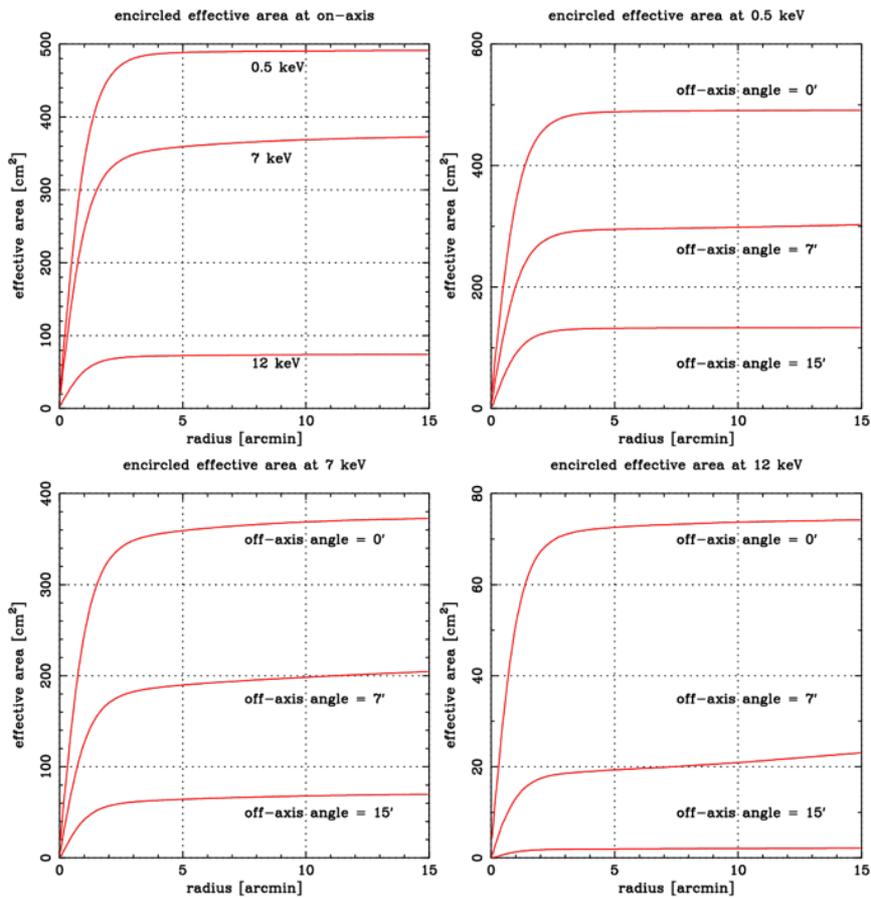

*Fig. 5* Encircled effective area, representing the sharpness of an image at several energies at on-axis and some off-axis angles.

# II-2. Resolve

## Basics

Resolve spectrometer is equipped with an X-ray microcalorimeter array of 6x6 pixels at the focus of the X-ray mirror assembly (XMA), which is capable of non-dispersive high-resolution ($\Delta E \leqq 7$ eV) spectroscopy and limited imaging of 3′ x 3′ field of view in the soft X-ray (0.3-12 keV) band with a large effective area of $\geqq 210$ cm$^2$.

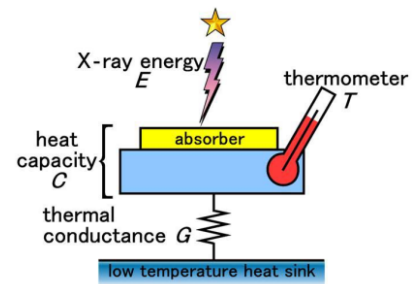

*Fig. 6* Concept of X-ray micro-calorimeter

## Technology

The detector measures the temperature rise upon each incident X-ray photon, achieving an unprecedented energy resolution. The thermometer is made of HgTe absorbers of an 8 μm thickness on micro-machined, ion-implanted Si wafer. The anti-coincidence Si semiconductor detector enables rejection of background events. The 50 mK operation temperature in the dewar is achieved by the cooling system comprised of adiabatic demagnetization refrigerators, He coolants, and mechanical coolers. The cooling system allows redundancy and cryogen-free operation. The filter wheel provides a suite of attenuation filters to enhance the dynamic range of the Resolve for bright sources. The modulated X-ray source (MXS) will be used for accurate energy-scale calibration in orbit.

## Performance

Since the Resolve is a non-dispersive spectrometer, it can be used to obtain high-resolution spectra of both point and extended sources (**Fig. 7**). The effective area of >200 cm$^2$ at the Fe K-band is considerably larger than any other high-resolution spectrometers. The imaging capability is limited as the HPD is comparable to the FoV (**Fig. 7**). Since the HPD is larger than the pixel size (~30″), there is a cross-contamination effect by the PSF. A simulation tool (e.g., heasim) estimates this effect for an observation target with multiple sources or an extended source with spatial variation in its spectrum.

The non-X-ray background (NXB) level after anti-coincidence screening should be lower than 2x10⁻³ counts/s/keV, which corresponds to 1.4 counts/(100 ks)/(7 eV). The major sources of the NXB are particles hitting the detector pixels, those hitting the support frame around the pixel array ("frame events"), and electrical crosstalk. The electrical crosstalk has a few 10-100 times smaller pulse height than its parental event, and therefore it is only a potential background component at very low energies in hard X-ray sources.

The Modulated X-ray Source (MXS) produces pulsed X-rays with ~a few ms pulse length and ~a few 100 ms pulse spacing, and provides fluorescent X-ray lines, mainly Cr Kα at ~5.4 keV and Cu Kα at 8.0 keV, for the gain tracking. When the MXS is used, MXS pulse-on time intervals are removed by a good-time interval (GTI) file computed by a dedicated tool. This leads to a small amount of loss of observing efficiency. Because MXS pulses have exponential tail, there are non-zero MXS events in pulse-off intervals. This contaminates GTI-screened data and contributes as another component of the background for Resolve. One of the pixels ("cal-pixel") is located outside the FoV and is illuminated by a Fe-55 source to monitor overall temporal gain drift of the detector. The cal-pixel, together with another Fe-55 source on the filter wheel, adds redundancy to the gain calibration.

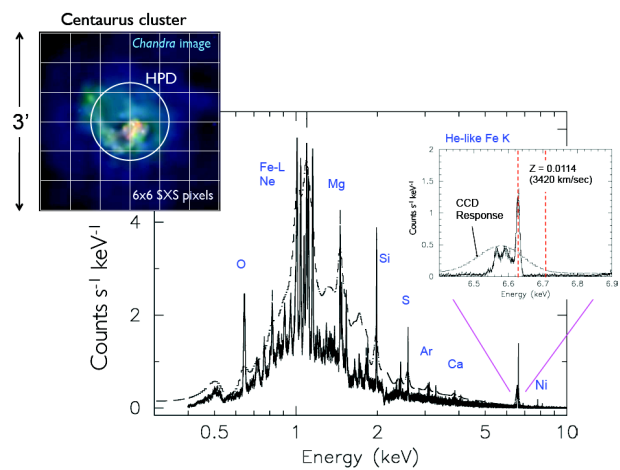

**Fig. 7** *Pixel layout (upper left panel). The array encompasses ~74% of photons of a point source at the center. Simulated spectrum of Centaurus cluster (bottom left).*

## Observation modes

Resolve has only one science observation mode. All X-ray events are graded based on intervals of event trigger times, which are determined by processing individual event pulse records (**Fig. 8**). Calorimeter resolution (ΔE≦7 eV) is guaranteed only for high-res (Hp) and mid-res primary (Mp) events. The digital processor (PSP) is capable of handling only up to 200/s ~ 120 mCrab ("PSP limit"; see **Fig. 9**). For bright sources, it is beneficial to intentionally reduce the incoming photons using filters (**Fig. 10**), offset pointings, or a combination of these.

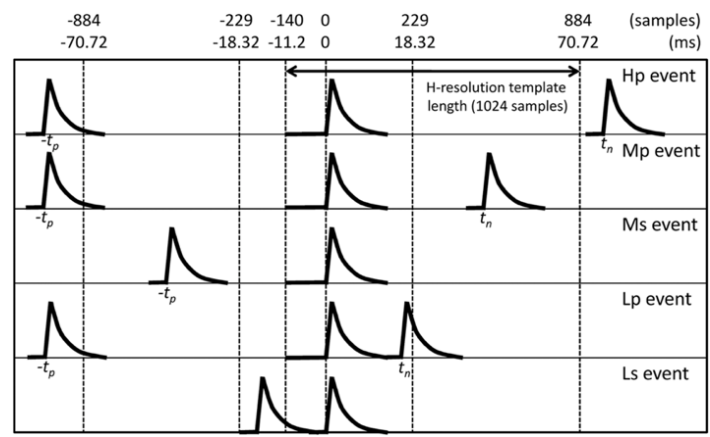

*Fig. 8* Grade definition. The figure illustrates spacings of event pulses (shown with triangular shapes) corresponding to the five event grades: high-res (Hp) if there is no event triggered within ±70.72 ms; mid-res (Mp and Ms) if there is no event within ±18.32 ms; else low-res (Lp and Ls). "p"=primary denotes the earlier event of an event pair, while "s"=secondary denotes the later one. The minimum separations to distinguish the grades (70.72 ms for mid-res vs. high-res and 18.32 ms for low-res vs. mid-res) are determined primarily by the length of per-event pulse record that is required by the PSP to perform the signal processing called the optimal filter.

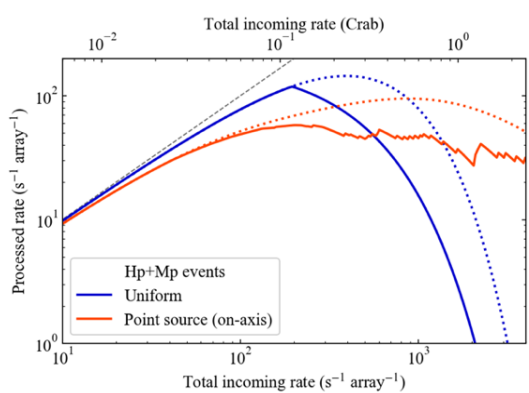

*Fig. 9* Incoming vs. processed rate for Hp+Mp events (solid). Because of the limitation in the CPU load ("PSP limit"), the processed rate is reduced from what would otherwise be observed (dotted). The calculations do not include the effect of electrical crosstalk.

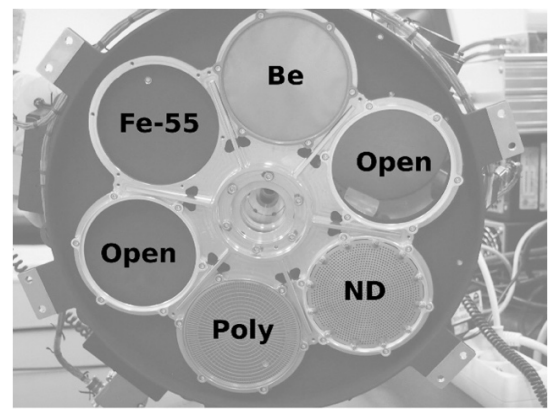

*Fig. 10* Filter selection. The filter wheel has three filters, two open positions, and one position with a radioactive Fe-55 calibration source. The three filters are Beryllium filter (Be), neutral-density filter (ND), and polyimide filter (Poly).

# II. Instruments
# II-3. Xtend

## Basics

Xtend is a telescope system which consists of an X-ray mirror assembly (XMA) and an X-ray CCD camera (SXI) [5]. It has an imaging-spectroscopic capability with a wide FoV (38x38 arcmin$^2$) and a medium energy resolution (E/ΔE~35@6 keV, based on the performance of Hitomi/SXI [7]) in the soft X-ray band (0.4-13 keV). It is a successor of Suzaku/XRT+XIS and Hitomi/SXT+SXI (**Table 3**).

*Table 3 Comparison of XRISM/Xtend and Suzaku/XRT+XIS.*

|  | No. of CCD | No. of telescope | Layout | Pixel scale | FoV [*1] (arcmin$^2$) | $A_{eff}$ [*2] (cm$^2$) | Grasp (cm$^2$ arcdeg$^2$) | Frame time (s) | Illumination | Type | Depl. layer (μm) |
|---|---|---|---|---|---|---|---|---|---|---|---|
| Xtend | 4 | 1 | 2x2 array | 1.77" | 38x38 | 374 356 | 54 | 4 | BI[*3] x4 | p-chan | 200 |
| XRT+XIS | 4 | 4 | 4 co-aligned | 1.04" | 18x18 | 694 797 | 35 | 8 | FI[*4] x3 BI[*3] x1 | n-chan | 80 (FI) 40 (BI) |

As the values of Xtend, the in-orbit performance of Hitomi/SXT+SXI is listed here. [*1]FoV for the 2x2 array for SXI. [*2]$A_{eff}$ is the on-axis effective area at 1 & 6 keV. The integration region is a circle with the radius of 1.8 arcmin. Four CCDs combined for XIS. [*3]Back-illuminated CCD. [*4]Front-illuminated CCD.

## Technology

The p-channel CCDs have a thick depletion layer to extend the hard-band coverage. The thick depletion layer will also reduce the background above ~7 keV compared to the same back-illuminated (BI) CCD in XIS. The BI CCDs have high quantum efficiency in the soft X-ray band. They are also more resistant than the FI CCDs to micro-meteorite hits. To decrease charge transfer inefficiency (CTI), a charge injection structure is equipped. It injects artificial charge to pixels periodically to fill traps which cause CTI. Mechanical coolers are used to keep the device temperature at -120 deg C. To suppress contamination accumulation on the CCD chips, a contamination blocking filter is placed on top of the hood. Onboard calibration sources ($^{55}$Fe) illuminate the corners of the CCDs. The design of Xtend is almost identical to that of Hitomi/SXT+SXI (**Fig. 11**) but some updates are introduced to solve the problems (e.g., optical light leak) found in the orbital operation of Hitomi [5].

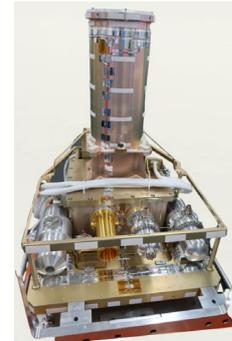

**Fig. 11** *External view of Xtend/SXI*

## Performance

The 2x2 CCD array covers a very large FoV. The on-axis position is 7.2' offset from the Xtend center. Xtend has a moderate energy resolution (**Fig. 12**). Like Suzaku/XRT+XIS and Hitomi/SXT+SXI (**Fig. 13**) [8], the background of Xtend is expected to be low and stable, benefiting from the low-earth satellite orbit.

## Observation modes

CCD clocking modes with different exposure times (4, 0.5, and 0.06 s) will be supported to mitigate CCD pile-up. The pile-up limits for each of the exposure time modes are about 1, 9, and 90 mCrab, respectively.

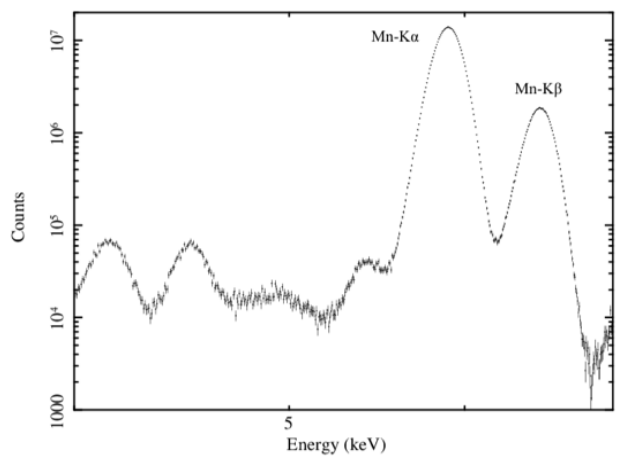

***Fig. 12*** *$^{55}$Fe Spectrum of a FM CCD chip in a ground calibration.*

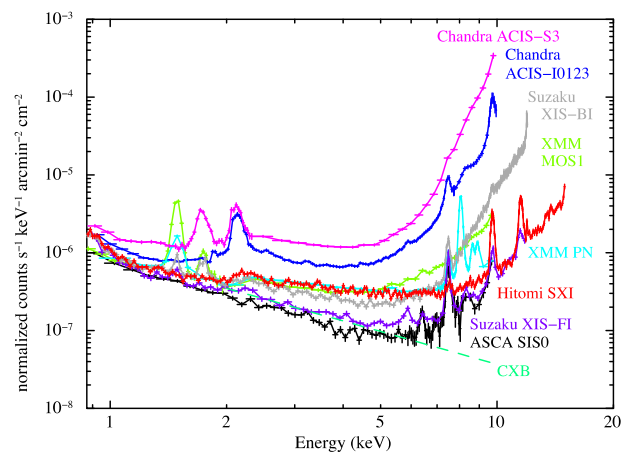

***Fig. 13*** *Comparison of the sky background of X-ray CCD cameras [8]*

## III. Figures of Merit
# III-1. Effective Area

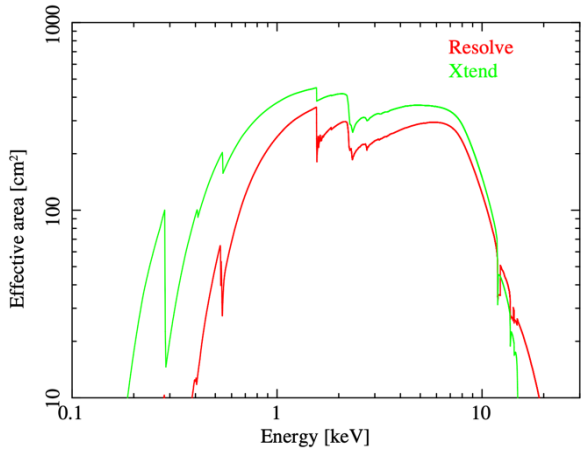

*Fig. 14* On-axis effective area of XRISM telescopes & detectors, Resolve (red) and Xtend (green). For Xtend, the in-orbit performance of Hitomi/SXT+SXI is plotted here.

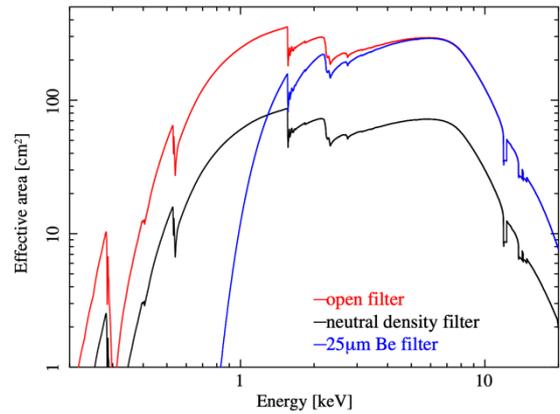

*Fig. 15* Resolve on-axis effective area with filter: the open filter (red), neutral density filter (black), 25um Be filter (blue).

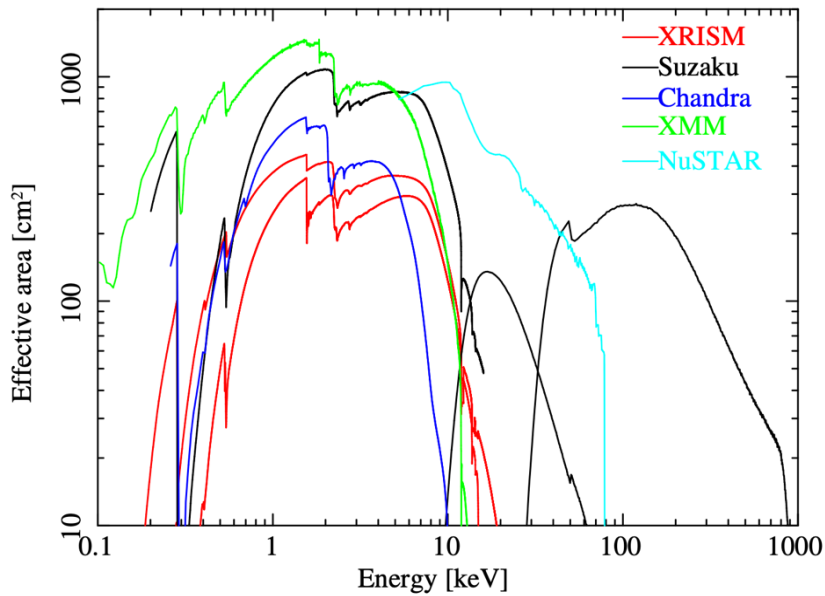

*Fig. 16* On-axis effective area of XRISM (Resolve, Xtend), Suzaku (XIS-BI+XIS-2FI, HXD), Chandra (ACIS-S), XMM-Newton/EPIC (PN+MOS1+MOS2), and NuSTAR.

# III-2. Energy Resolution

(i) Strong line: FOM ~ √A

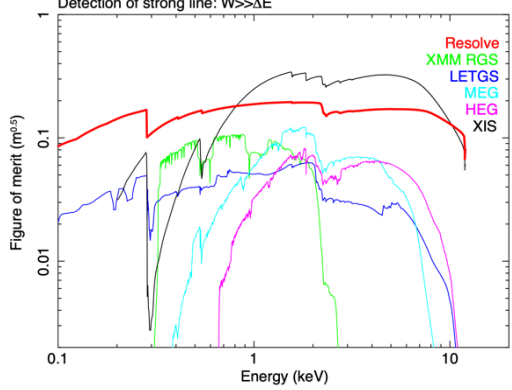

(ii) Weak line: FOM ~ √(A/ΔE)

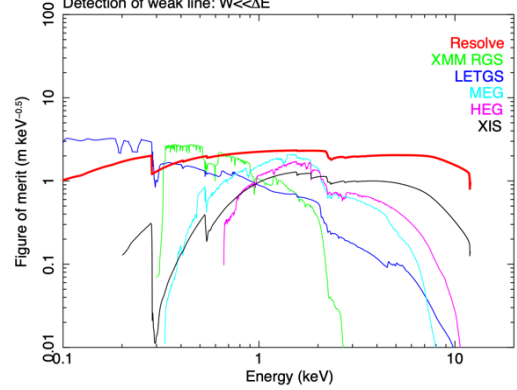

(iii) Strong line: FOM ~ √(AE$^2$/ΔE$^2$)

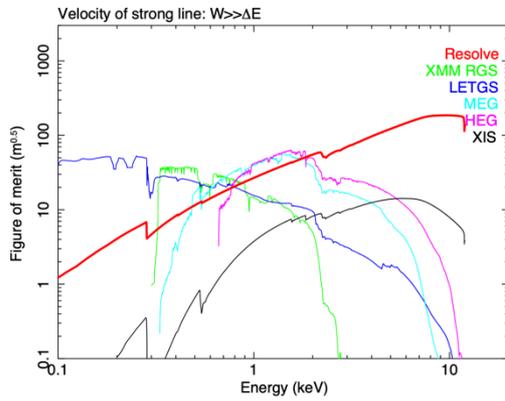

(iv) Weak line: FOM ~ √(AE$^2$/ΔE$^3$)

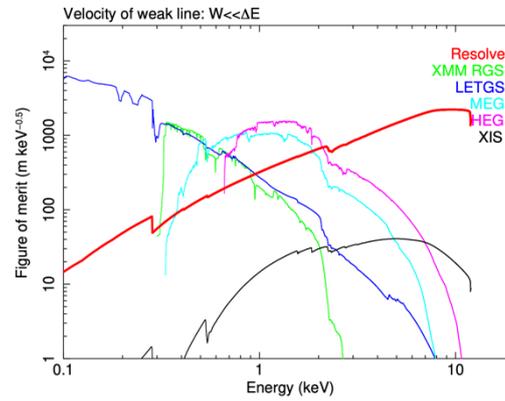

(v) Strong line: FOM ~ √(AE$^4$/ΔE$^4$)

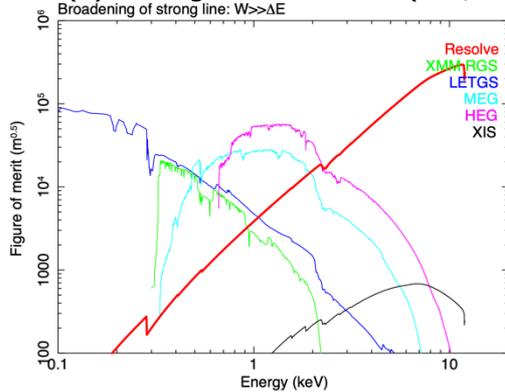

(vi) Weak line: FOM ~ √(AE$^4$/ΔE$^5$)

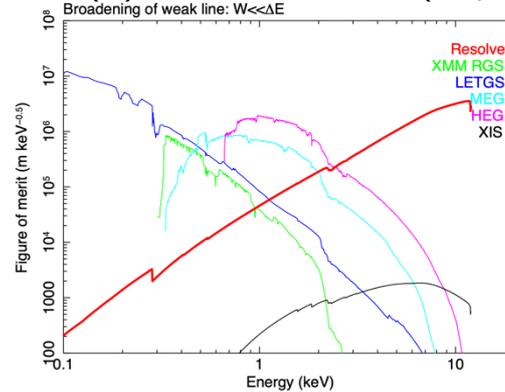

*Fig. 17* Figures of merit (FOM) for (i) detection of strong line, (ii) detection of weak line, (iii) velocity of strong line, (iv) velocity of weak line, (v) broadening of strong line, (vi) broadening of weak line. Strong lines are defined here as lines with equivalent width W larger than the instrumental resolution ΔE, weak lines have W smaller than ΔE.

## III. Figures of Merit
# III-3. FoV and Grasp

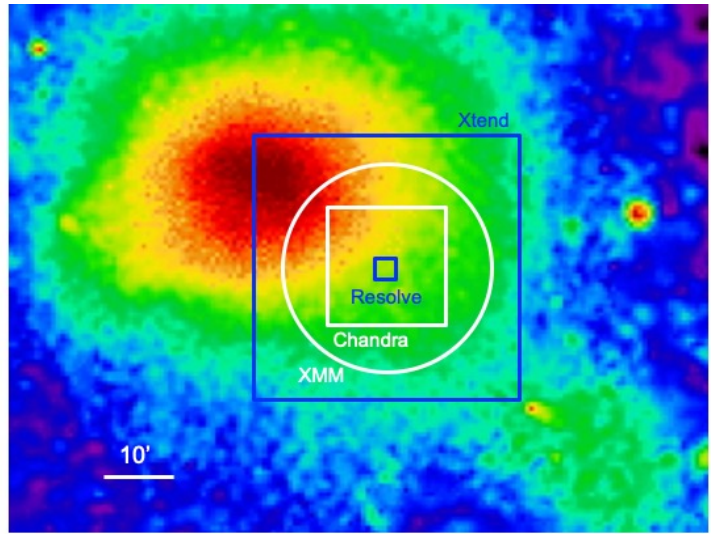

*Fig. 18* *Field of view of the XRISM instruments, Resolve and Xtend (the blue boxes). Chandra ACIS-I and XMM are also shown for comparison. The background image is the Coma cluster taken with ROSAT (credit: ROSAT/MPE/S. L. Snowden).*

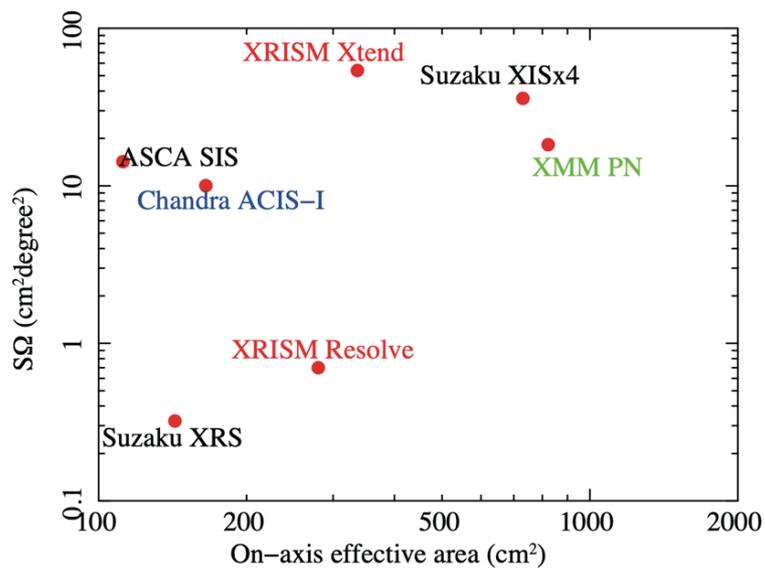

*Fig. 19* *Grasp vs on-axis effective area at 7 keV of Resolve and Xtend. Suzaku XIS, Chandra ACIS-I, XMM PN are also shown for comparison. The grasp is defined as the product of field of view and effective area.*

# IV-1. Black Holes

XRISM provides an unprecedented view of the motions and extreme physical conditions of matter near the event horizon. This will help us unravel how black holes grow by accreting gas and simultaneously shape their environments via the intense radiation fields and powerful, sometimes relativistic outflows that accompany this accretion.

Gaseous winds driven from black hole disks can carry away a substantial fraction of the gas that would otherwise accrete onto the central black hole. Such winds are hot, and most easily detected in the Fe K band via absorption lines. As shown in **Fig. 20**, the superior resolution of Resolve in the Fe K band enables the unambiguous detection of weak and narrow lines from a wind. We will be able to use these to precisely determine the radius at which the wind is launched, and the mass outflow rate carried by it. This will give us strong constraints on the driving mechanism of the wind and its feedback on the accretion flow.

XRISM will also give us key observations to investigate the feeding and feedback mechanisms in supermassive black holes (SMBHs) at galactic centers, and thereby the origin of SMBH-galaxy coevolution. Cold matter surrounding an active galactic nucleus (AGN), called the "torus", serves as a mass reservoir linking the SMBH and host galaxy. Outflows of warm photoionized gas by the central radiation carry significant mass, momentum, and energy outward, which have strong impacts on its environment.

XRISM will provide us with an unbiased view of mass distribution and kinetics of surrounding matter, through observations of fluorescence lines from the cold torus and absorption/emission lines from highly ionized outflows. As shown in **Fig. 21**, Resolve enables us to separate and characterize many emission lines from different locations, which tell us the physical origins of these phenomena.

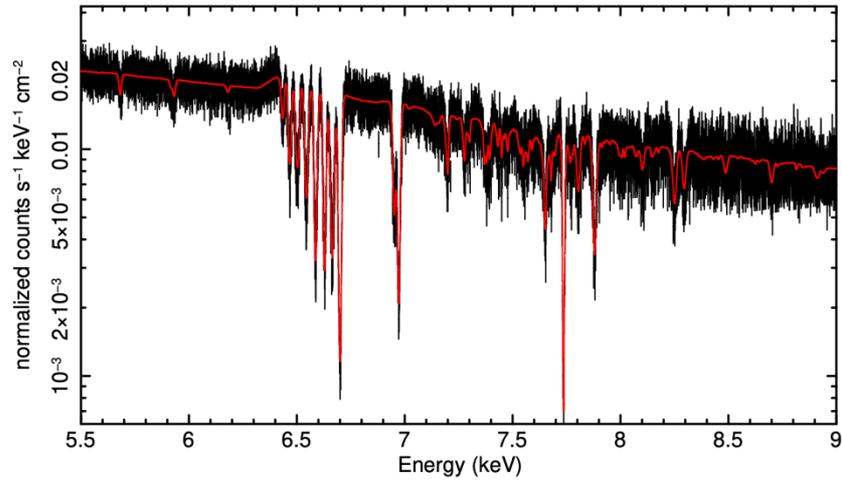

***Fig. 20*** *A simulated spectrum of the black hole GRS 1915+105 with a 40 ks Resolve observation. The model is based on a prior Chandra observation (Miller et al. 2020).*

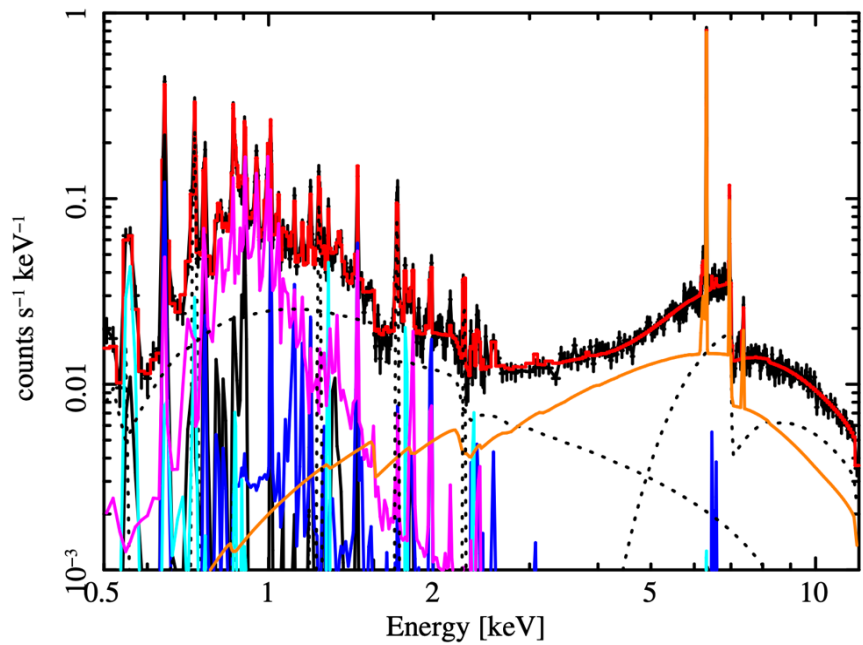

***Fig. 21*** *A simulated spectrum of the obscured AGN Mrk 3 with a 200 ks Resolve observation. The model is based on a prior Suzaku observation (Awaki et al. 2008).*

# IV. Case Studies

# IV-2. Supernova Remnants

The high-resolution spatially-resolved X-ray spectroscopy provided by Resolve will be particularly ground-breaking for supernova remnants (SNRs) because they are extended objects with rich emission-line spectra from a wide range of different elements (carbon through nickel). The following examples highlight some of the topics that XRISM will address.

**(1) Understanding the progenitors of various kinds of SNRs**

Low-abundance elements such as Ti, Ni, Cr, Mn are the key elements to determine the physical condition of the stellar core at the moment of explosion. XRISM will resolve emission lines from all four of these elements and measure their total abundances, placing constraints on how the original supernova explosion progressed (**Fig. 22**). The velocity of the rapidly expanding supernova ejecta, as well as the circumstellar and interstellar gas it sweeps up will also be measured by XRISM via Doppler shifts. Such velocity measurements reveal how SNRs evolve, based on their age and the detailed properties of the explosion, the ejecta, and ambient medium.

**(2) Understanding the physics of collisionless shocks in SNRs**

Collective interactions between particles and electromagnetic fields are responsible for shock formation (collisionless shocks) and energy redistribution. XRISM will resolve this heating process, as it impacts electrons, protons, and heavier elements. Shocks accelerate some particles into the TeV range, believed to be the origin of Galactic cosmic rays, and these particles extract energy from the shocks. XRISM will measure the fraction of the shock's energy that goes into accelerating particles and the particle acceleration efficiency.

**(3) Understanding Radiative processes of hot plasmas**

There are many details in the X-ray spectra of hot plasmas that remain unexplored due to the limited capabilities of the current generation of X-ray spectrometers. XRISM will show us the first precise measurements of the plasma in various states, ionizing or recombining, and various effects such as resonance scattering and charge exchange (**Fig. 23**).

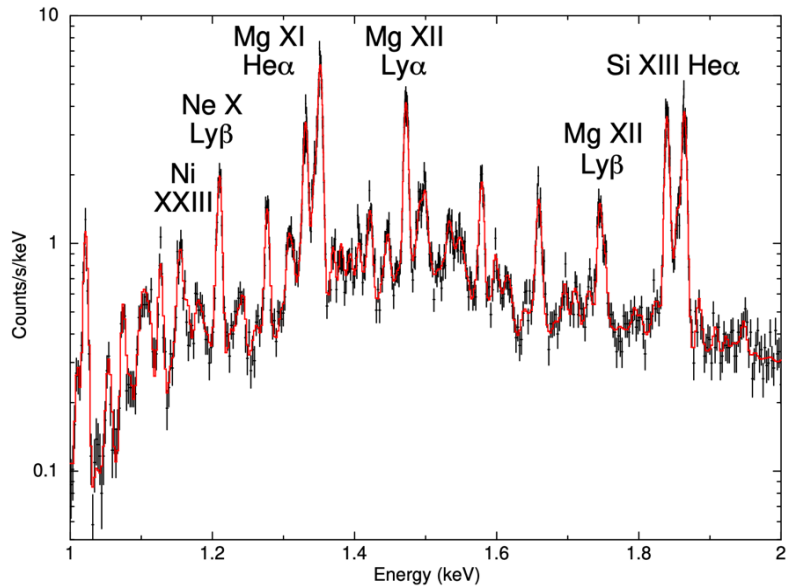

*Fig. 22* *100ks simulation of western half of 3C397. Several lines via Ni and Fe L-shell emission are seen for example at 1.045,1.156 keV for Ni and 1.01 keV for Fe, respectively as well as peaks from K-shell transition of Si (1.84 keV) and Mg (1.33 keV).*

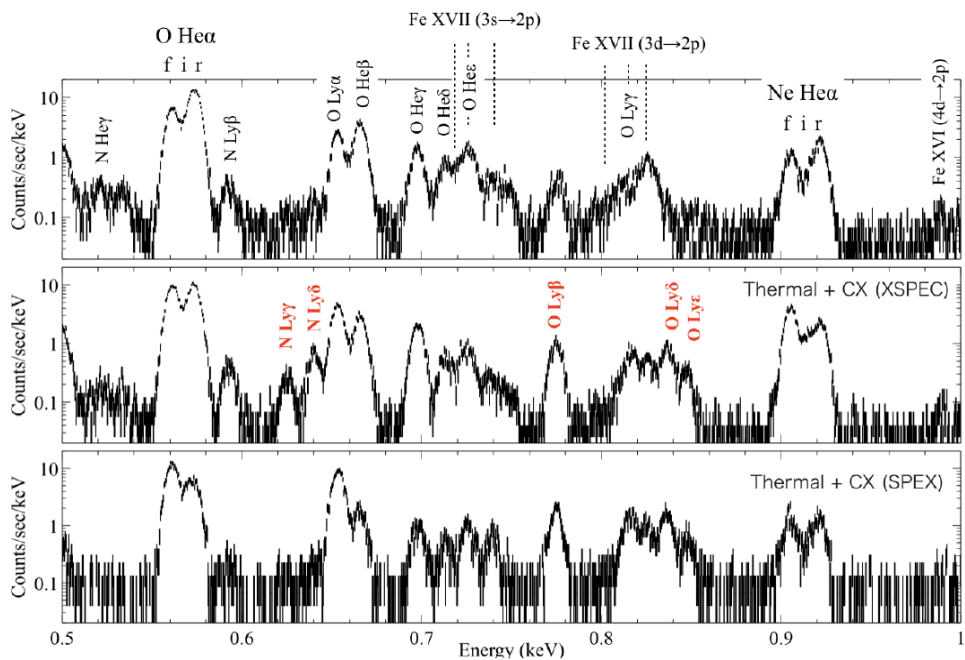

*Fig. 23* *30-ks simulated Resolve spectra of a shell region of the Cygnus Loop. Top: Assumed model is a typical thermal plasma with electron temperature of 0.2 keV. Middle: Same as the top panel but added a charge-exchange model in XSPEC. Bottom: Same as the middle panel but simulated in SPEX with similar model parameters.*

**17**

# IV-3. Clusters of Galaxies

The dominant baryonic component of galaxy clusters is in a form of hot intergalactic gas. Shocks produced by violent mergers of clusters and matter accretion, feedback from Active Galactic Nuclei (AGNs) heat the gas to very high temperatures. With its calorimetric energy resolution, the *Resolve* will detect individual spectral lines emitted by the hot gas, allowing the measurements of turbulence and gas flows generated during these events. The low-background *Xtend* will map the gas thermal state in cluster outskirts. These measurements are crucial for understanding the evolution of the largest objects in the universe and precise cluster cosmology.

**Feedback from supermassive black holes**

In nearby bright galaxy clusters, like the Virgo/M87 cluster, XRISM will spatially and spectrally resolve the interaction between AGN and intracluster medium, allowing the studies of the mixing of different gas phases and metals and probing gas dynamics close to the supermassive black hole. White squares in **Fig. 24** show the Resolve field of view. Simulated 100-ks spectrum shows that the kinematics of two different gas components can be measured using the Fe-L line complex.

**Non-thermal pressure support for precise cluster cosmology**

XRISM will measure turbulence contribution to the total pressure in galaxy clusters significantly improving their mass measurements that are crucial for precise cluster cosmology. Simulated *Resolve* spectrum from the region around $r_{2500}$ in A2029 shows strong Fe lines broadened by turbulence (**Fig. 25**).

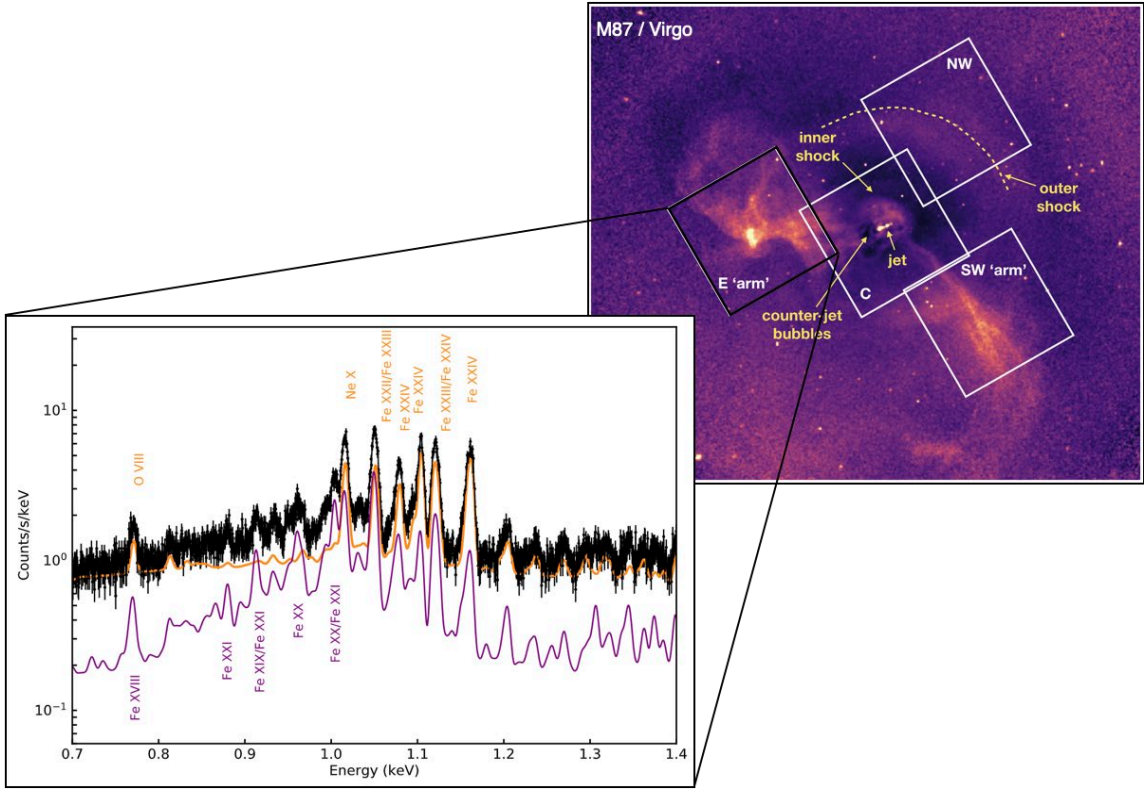

***Fig. 24*** *Simulated 100-ks Resolve spectrum of M87 (black). Contributions from different thermal components to the Fe-L line complex are shown (orange: hot atmosphere, purple: uplifting gas,).*

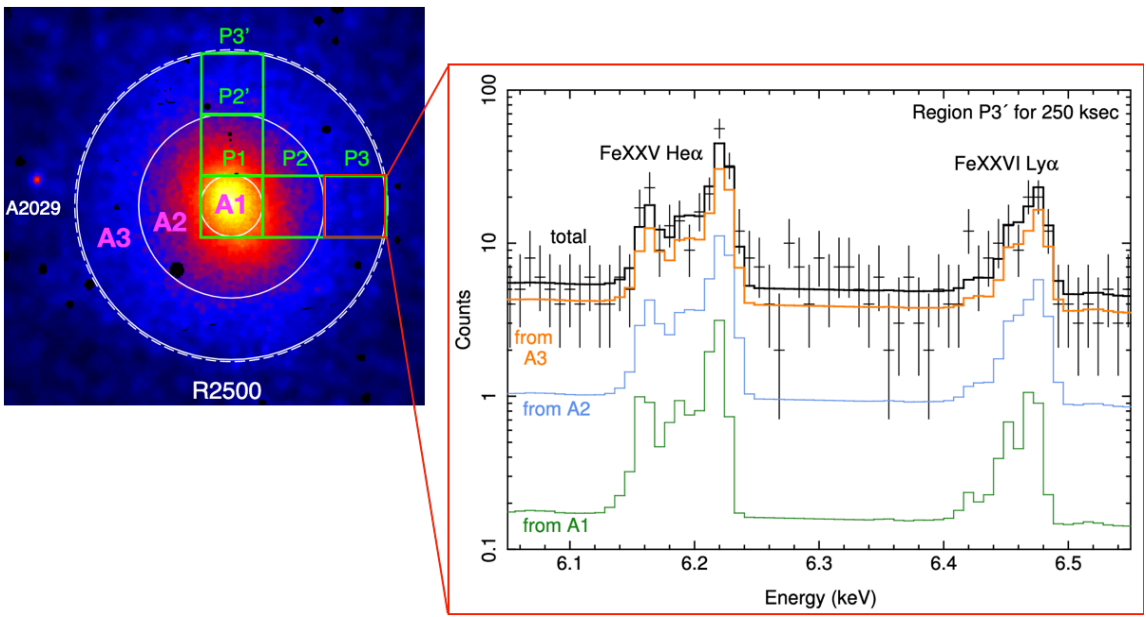

***Fig. 25*** *Simulated 250-ks Resolve spectrum of the outermost A2029 pointing, P3′ (black). Contributions from each annulus are shown in different colors (green: A1, blue: A2, orange: A3), suggesting that correction for the contamination from the bright core is necessary for the accurate turbulence measurement.*

**19**

# Further Information

**Web page**

- XRISM <http://xrism.isas.jaxa.jp>

**References**

- XRISM

  [1] Tashiro et al. 2018, SPIE, 10699, id. 106992

- Resolve

  [2] Kelly et al. 2016, SPIE, 9905, id. 99050V

  [3] Sato et al. 2016, JATIS 2d4001S

  [4] Kilbourne et al. 2018, JATIS 4a1214K

  For subsystems and calibration, a comprehensive list is available: https://heasarc.gsfc.nasa.gov/docs/hitomi/about/paper_list.html#jatis

- Xtend

  [5] Hayashida et al. 2018, SPIE, 10699, 1069923, SPIE10699

  [6] Nakajima, Hitomi Collaboration, 2017, NIMPA, 873, 16

  [7] Tanaka et al. 2018, JATIS, 4, 011211

  [8] Nakajima et al. 2018, PASJ, 70, 21

- Hitomi White Papers

  https://heasarc.gsfc.nasa.gov/docs/hitomi/about/paper_list.html#whitep

**Point of Contact**

- Questions and comments to the document
    - SOC-PVO <Z-xrism-soc-qref@ml.jaxa.jp>

**Acknowledgement**

A. Bamba, M. Diaz Trigo, M. Guainazzi, J. Kaastra, Y. Fujita, K. Matsushita, R. Mushotzky, H. Nakajima, J.-U. Ness, M. Sawada, H. Uchida, Y. Uchida, Y. Ueda, H. Yamaguchi, I. Zhuravleva

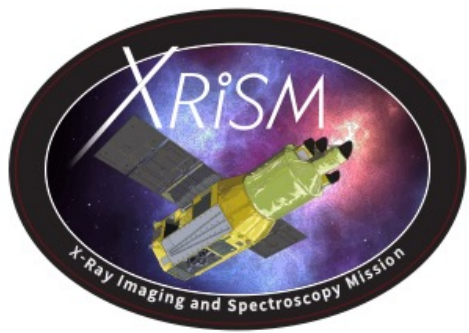